# A Modified Thomas-Fermi Approximation With Applications

**Gregory C. Dente**

*GCD Associates, 2100 Alvarado NE, Albuquerque, NM 87110*

**Abstract**

In order to obtain a reasonably accurate and easily implemented approach to many-electron calculations, we will develop a new Density Functional Theory (DFT). Specifically, we derive an approximation to electron density, the first term of which is the Thomas-Fermi density, while the remaining terms substantially correct the density near the nucleus. As a first application, this new result allows us to accurately calculate the details of the self-consistent ion cores, as well as the ionization potentials for the outer s-orbital bound to the closed-shell ion core of the Group III, IV and V elements. Next, we demonstrate that the new DFT allows us to separate closed-shell core electron densities from valence electron densities. When we calculate the valence kinetic energy density, we show that it separates into two terms: the first exactly cancels the potential energy due to the ion core in the core region; the second represents the residual kinetic energy density resulting from the envelopes of the valence electron orbitals. This kinetic energy cancellation in the core region and the residual valence kinetic energy term allow us to write a functional for the total valence energy dependant only on the valence density. This equation provides the starting point for a large number of electronic structure calculations. Here, we use it to calculate the band structures of several Group IV and Group III-V semiconductors.

## I. Introduction

Calculating the properties of atoms, molecules and solids has been one of the primary objectives of physics for the last century. Certainly, by 1930, the machinery of quantum mechanics was well-understood and spectacularly successful when applied to one- and two- electron systems. However, as researchers began to tackle other many-electron problems, the calculations quickly became complicated and unwieldy, leading P.A.M. Dirac to famously say: *"The fundamental laws necessary for the mathematical treatment of a large part of physics and the whole of chemistry are thus completely known, and the difficulty lies only in the fact that application of these laws leads to equations that are too complex to be solved."*[1,2] One powerful approach, variational calculations based on determinant wave functions, the Slater Determinant, led to a set of N-coupled integral-differential equations for N single-electron orbitals.[3,4] These nonlinear Hartree-Fock equations then had to be solved self-consistently. Although many solutions have been obtained over the last eighty years, when the number of electrons became large, these procedures proved difficult. In fact, for large N, one could question the very practicality of an antisymmetric N-electron wave function that is a function of 3N coordinate variables.[1]

As an alternative to solving for an N-electron wave function, researchers also developed methods that dealt directly with the electron density. These density-functional theories, DFT, can be derived from, or at least motivated from, the N-electron wave equation.[1] The earliest example of a DFT was developed in the late 1920s; this is the Thomas-Fermi model, one of the earliest schemes for calculating the N-electron problem while enforcing the Pauli exclusion principle and wave-particle duality. In this Fermi gas- motivated model, the local electron density is related to the Fermi momentum as

$$\rho(\vec{r}) = 2 \cdot \frac{4\pi P_F^3(\vec{r})}{3(2\pi\hbar)^3} = \frac{(2m(F-V(\vec{r})))^{3/2}}{3\pi^2 \hbar^3} \quad , \tag{1}$$

in which the Fermi momentum, $P_F$, of the most energetic electrons is specified by the Fermi energy, F, and the local potential, V. When the Poisson potential, as well as the exchange/correlation components of the potential, could be determined by the density, a self-consistent solution was then possible.[3,4] Unfortunately, Thomas-Fermi has always been considered a crude approximation, not accurate enough for quantitative chemistry or material science calculations.[1] This paper will demonstrate that a new density expression, one that bears a resemblance to Thomas-Fermi, can consistently yield highly accurate solutions. Additionally, the new method leads to two remarkably helpful developments. First, we can readily separate valence electron densities from core electron densities. Second, we can show that the valence kinetic energy density can be separated into a term that exactly cancels the potential, due to the nucleus and closed-shell core electrons in the core region, while the remaining term can be interpreted as a residual kinetic energy density generated by the envelopes of the valence orbitals. This type of kinetic energy cancellation, based on an envelope function approximation for the valence orbitals, is a critical element for making calculations tractable. Once the problem has been reduced to one in which the high-potential gradients near the nuclei have been removed from the valence electron equations, the remaining low-spatial-frequency (LSF) phenomena, determined only by valence electron densities, are far easier to calculate while retaining substantial accuracy for electronic structure calculations.

In Section II, supplemented by Appendix A, we will derive an improved form of the electron density. In the following Section III, we will present an application which allows us to calculate a set of ionization potentials for the ionized Group III, IV and V elements; in all cases, we will give results for the energy of the outer s-orbital bound to the closed-shell ion core. Next, Section IV presents a method for separating valence electron densities from the closed-shell core densities. A rearrangement of the valence electrons' kinetic energy then shows cancellation of the core potential with the dominant part of the kinetic energy density of the valence electrons; a residual valence kinetic energy density remains. Section V implements these effects by developing an equation for the total energy of the valence electrons in which a much weaker effective potential, $V_{eff}$, replaces the strong ion core potential. Appendix B, supplementing Sections IV and V, develops the valence orbital origins of the core potential cancellation by factorizing the orbitals into a radially oscillating function times slowly-varying envelopes; an envelope function approximation to the orbital energy density then leads to the weaker effective potential. In Section VI, we present a method for setting the parameters of the effective potential, particularly the core radius of the closed-shell ion. In Section VII, we bring all the pieces together to present our band structures resulting from the self-consistent valence density and potential on the zinc-blende lattice. We will give accurate band structures for the very technologically important materials, silicon and gallium arsenide, as well as show results for many other semiconductor materials. Section VIII contains our conclusions, along with a brief discussion relating these methods to the rich history of electronic structure calculations.

Judicious use of approximations has always defined the art of physics applied to materials science and chemistry, especially for elements in the higher-numbered periods of the periodic table. The many-electron problem in quantum mechanics has traditionally been a domain showered with approximations. In this paper, we will use the Wentzel-Kramers-Brillouin (WKB) approximation for orbitals, invoke envelope function approximations as needed, replace sums over orbitals with integrals over energy ranges and finally, only offer estimates for the critical closed-shell ion core radius parameters. Despite the approximations and intuitive insights, the final working equations, Eqs. (2-4) for atomic physics problems, and Eqs. (22-23) for calculations on the outer valence electrons in multiple atom problems, appear to be accurate and easy to implement, providing self-consistent electron densities and potentials for a large class of many-electron calculations.

## II. The Density Expression: Lowest orbital corrections

It is well-known that the Thomas-Fermi density of Eq. (1) gives an infinite density to the electrons near the nucleus.[4,5] This then leads to errors in total binding energy, which have been studied and substantially corrected.[5] Here, we advocate a different approach and look for a revised formula for the electron density. Appendix A gives the details of our new derivation done two different ways for systems in which there are equal densities of spin-up and spin-down electrons. The final result is

$$\rho(\vec{r}) = \frac{(2m(F-V(\vec{r})))^{3/2}}{3\pi^2 \hbar^3} - \frac{(2m(E_0-V(\vec{r})))^{3/2}}{3\pi^2 \hbar^3} + |\Psi_0(\vec{r})|^2 ,$$
$$= f(F-V(\vec{r})) - f(E_0-V(\vec{r})) + |\Psi_0(\vec{r})|^2$$

(2)

in which we define the density function for potential $V(\vec{r})$ up to Fermi level, $F$, as

$f(F-V(\vec{r})) \equiv \frac{(2m(F-V(\vec{r})))^{3/2}}{3\pi^2 \hbar^3}$ ; note that this density function is defined to be zero when the argument is less than or equal to zero, $F - V(\vec{r}) \leq 0$. Finally, $\Psi_0(\vec{r})$ is the orbital at the lowest energy, $E_0$.

Eq. (2) contains the Thomas-Fermi density as the first term on the right-hand side, but then differs from the standard result in several important ways. First, the modified density results from the WKB approximation to the electron orbital envelopes, as well as a direct application of the Euler-Maclaurin formula, converting a sum over orbital densities to an integral over energy;[6,7] additionally, Appendix A then bolsters the modified density result with an alternative, operator algebra-based derivation. These derivations, the special handling of the lowest-energy orbital terms and the resulting density modifications suggest a higher level of accuracy than is present in the standard Thomas-Fermi result based on a Fermi gas approximation.[3,4] Second, for atomic problems, this density remains finite at the nucleus, completely curing the divergence of the Thomas-Fermi density. Third, when the Poisson potential and the exchange/correlation components of the potential can both be approximated from the electron density, a self-consistent solution using Eq. (2) gives consistently accurate results for a variety of atomic structure problems. Fourth, the improved density expression also leads to a method for separating closed-shell core and valence electron densities. In the following sections, we will demonstrate these features with applications to both ionization potentials and band structure calculations.

## III. Predicting Ionization Potentials for Closed-Shell Ions

As a first application of the new density expression, we will investigate its accuracy in closed-shell ion cores. Specifically, we will use it to calculate the third ionization potential of the Group III elements, the fourth ionization potential of the Group IV elements and finally, the fifth ionization potential of the Group V elements. In all cases, we will calculate the new density, $\rho(r)$, and self-consistent potential, $V(r)$, for the closed-shell ion, and then solve the radial wave equation for the 'ns'-orbital electron energy and wave function using this self-consistent potential. This approach neglects any influence of the lowest-energy valence electron on the closed-shell ion core and is, therefore, approximate. However, the final results are in good agreement with the measured ionization potentials.

We will detail our calculation for $Si$ with atomic number $Z=14$. We need to find the energy, $I(4)$, required to remove the outer electron from $Si^{+3}$, so that
$Si^{+3} + I(4) \rightarrow Si^{+4} + e$.

First, we calculate the density and potential for the ten core electrons in $Si^{+4}$. The total potential is given as the sum of the electrostatic and exchange/correlation parts

$$V = V_P + V_{exc} \quad , \tag{3}$$

in which $V_P$ satisfies the Poisson equation in the radial coordinate as

$$\nabla^2 \left( V_P + \frac{Z e^2}{r} \right) = \frac{1}{r} \frac{d^2}{dr^2} \left( r \cdot \left( V_P + \frac{Z e^2}{r} \right) \right) = -4\pi e^2 \rho(r) \quad . \tag{4}$$

The electron density on the right-hand-side is the new DFT given in Eq. (2).

To calculate the density, we must estimate the lowest orbital, $\Psi_0(r)$, and the energy, $E_0$. Here, we are guided by the variational form for a two-electron singlet wave function as

$$\Psi_0(r_1) \cdot \Psi_0(r_2) = \frac{\alpha^3}{\pi \cdot a_0^3} \exp(-\alpha(r_1 + r_2)/a_0) \quad . \tag{5}$$

When we minimize the energy of a two-electron atom with charge Z on the nucleus, we find a minimum at $\alpha = Z - 5/16$ and $E_0 = -(Z-5/16)^2$ Rydbergs. This is the standard shielding result for the inner two electrons while neglecting all others.[4] Actually, many results for valence electrons are relatively insensitive to the lowest-orbital estimate, but this approximation for the lowest orbital leads to excellent results and, in addition, the exponential form near the nucleus seems correct.

The exchange/correlation parts, $V_{exc}$, are approximated in the local density approximation (LDA) as the derivative of the exchange/correlation energy density,

$$V_{exc} \equiv \frac{dU_{exc}}{d\rho} = -(3/\pi)^{1/3} \cdot e^2 \rho(r)^{1/3} \quad , \tag{6}$$

in which we have neglected correlation effects.[1,3,11] This functional form can be motivated from the "Fermi hole" that each electron forms and carries with it in the presence of parallel spin electrons. The idea of this local density approximation for exchange is due to Slater, who derived an exchange potential that was 3/2 times Eq. (6). The correction was obtained by Kohn and Sham.[2,4] Actually, many early calculations correctly gave the average exchange energy density, $U_{exc}(\rho)$.[4,5,11] The potential then results from differentiating the energy with respect to density; this is the standard Kohn and Sham prescription.[1,9,15]

We solve Eqs. (2) through (6) with an iterative procedure. First, from previous values of $V_{exc}$, we use a predictor-corrector integrator to solve Eq. (4) for the updated Poisson potential on a radial grid. We then adjust the Fermi level and repeat the integration until the volume integral of the density converges to $Z-v=10$ for $Si^{+4}$, in which $v=4$ is equal to the valence for silicon. Next, we use these converged density values to find new estimates for $V_{exc}$. We then return to the first step and iterate to overall convergence. The final output of this procedure includes the electron density, $\rho$, the potentials, $V_P$ and $V_{exc}$, of the closed-shell ion core, as well as the radius, $R_{ion}$, defined by the radial value at which the self-consistent electron density falls to zero.

We estimate the ion core potential and continuously connect it to the outer Coulomb potential as

$$V_c(r) = V_P(r) - V_P(R_{ion}) + \kappa \cdot V_{exc}(r) - \frac{v \cdot e^2}{R_{ion}} \qquad r \leq R_{ion}$$
$$V_c(r) = -\frac{v \cdot e^2}{r} \qquad r \geq R_{ion}$$
. (7)

This should approximate the ion core potential sensed by the lowest-energy valence electron; however, it neglects the influence of the valence electron back on the core. Also, we have included a constant factor, $\kappa$, in order to make adjustments in the strength of valence electron interaction with the ion core LDA exchange potential. $\kappa$ can be simply treated as an adjustable parameter; however, we can make a factor less than unity plausible: In the exchange contribution to the Hartree-Fock eigenvalues for plane-wave orbitals, we encounter a factor multiplying the average exchange potential Eq. (6). This factor is

$$\kappa = 1 + \frac{1-\eta^2}{2\eta} \log\left|\frac{1+\eta}{1-\eta}\right| \quad , \tag{8}$$

in which $\eta$ is the local ratio of the valence electron momentum to the Fermi momentum defined by the ion core electron density.[3,4] In keeping with the plane-wave approximation, since a valence orbital has higher energy than the core electrons, it becomes clear that $\eta > 1$ and, therefore, $\kappa < 1$ for a valence electron in the core region. We use an average constant $\kappa$ in Eq. (7) to roughly mimic this behavior.

Next, we calculate the energy eigenvalue, $\varepsilon$, and orbital eigenfunction, $\psi \equiv \phi / r$, of the s-orbital bound to the closed-shell ion core potential by using a predictor-corrector integrator on the radial wave equation ($l = 0$):

$$\frac{d^2 \phi}{dr^2} = \left( \frac{l(l+1)}{r^2} + \frac{2m}{\hbar^2} V_c(r) - \frac{2m}{\hbar^2} \varepsilon \right) \phi \quad . \tag{9}$$

We iterate the radial integrations and adjust the energy eigenvalue until the 'ns'-orbital eigenfunction converges at large radii.

Table I presents our s-orbital eigenvalues for the Group III, IV and V closed-shell ions. Since Eq. (8) suggests that the valence electron might respond to less of the core exchange potential, we show results for two values of $\kappa = .5$ and $\kappa = 1$. We also show the experimental results for these ionization potentials. With the exception of the fourth period ions, $Ga, Ge, As$, the values calculated for the range $.5 \leq \kappa \leq 1$ bracket the experimental results. We would obtain better fourth period results with $\kappa$ slightly less than .5. As expected, the Group IV 'ns'-orbital eigenfunctions for $C^{+3}, Si^{+3}, Ge^{+3}, Sn^{+3}$ and $Pb^{+3}$ show 2s, 3s, 4s, 5s and 6s character respectively. Perhaps, if we used the more exact non-local form of exchange potential in the radial wave equation, or allowed the core to be perturbed by the valence electron, we could further improve these LDA results.[4] Despite these approximations, it appears that the new DFT provides reasonably accurate descriptions of the closed-shell ion cores for Groups III, IV and V of the periodic table. These results will be used later in Section VI.

## IV. Separating Core from Valence Electrons: Kinetic and Potential Cancellation

With accurate core results established, we can begin to consider the dynamics of the outer valence electrons, the electrons that are the critical players in the chemical bond. For each constituent atom, we will always equate the valence to the number of electrons outside the closed-shell core. One extremely convenient feature of the new density expression is the simple separation of valence electrons from core electrons. Consider rewriting Eq. (2) as

$$\rho(\vec{r}) = f(F - V(\vec{r})) - f(F_c - V(\vec{r})) \\ + f(F_c - V(\vec{r})) - f(E_0 - V(\vec{r})) + |\Psi_0(\vec{r})|^2 \quad . \quad (10)$$

In this form, we can tentatively identify the two terms in the first line as the valence density, $\rho_v$, and the three terms in the second line as the closed-shell core density, $\rho_c$. Here, the additional parameter is the Fermi level setting for the core, $F_c$. In applications it is adjusted, as in Section III, to fix the number of closed-shell core electrons. This valence density can also be evaluated as the derivative, $\frac{df}{dE}$, the change in electron density with respect to energy, integrated over the valence energy range, $F_c \leq E \leq F$.

We can readily calculate the kinetic energy density of these valence electrons. The final result involves the derivative, $\frac{df}{dE}$, times the kinetic energy factor, $(E - V(\vec{r}))$. This integrand is then integrated over the range of valence energies as

$$t_v(\vec{r}) \equiv \int_{F_c}^{F} dE (E - V) \cdot \frac{df(E - V)}{dE}$$
$$= \frac{3}{5}(F_c + \delta F - V) f(F_c + \delta F - V) - \frac{3}{5}(F_c - V) f(F_c - V) \quad r \leq r_c \quad , \quad (11)$$
$$= \frac{3}{5}(F_c + \delta F - V) f(F_c + \delta F - V) \quad r > r_c$$

in which we redefine the upper Fermi level as $F \equiv F_c + \delta F$, so that the valence electrons occupy an energy range, $\delta F$. Also, we introduce a new parameter, the core radius, defined by the equation $F_c - V(r_c) \equiv 0$. In the WKB approximation to the core electron orbitals, this core radius marks the boundary between the classically attainable, oscillatory core region and the classically unattainable, damped region.[6] We will have more to say about $r_c$ in the next sections.

This valence kinetic energy density can be rearranged in a suggestive and useful way. Consider the exact rewrite of Eq.(11) as

$$t_v(\vec{r}) = (-V_c)\rho_v + \int_0^{\delta F} ds(s + F_c - V_v)\frac{df(s + F_c - V)}{ds} \qquad r \leq r_c$$
$$= \int_0^{\delta F} ds(s + F_c - V)\frac{df(s + F_c - V)}{ds} \qquad r > r_c \qquad , \quad (12)$$

in which $V_c$ is the potential (direct, exchange and correlation) due to the ion core, $V_v$ is the potential due to the valence electrons and $V = V_c + V_v$. The first term in this kinetic energy density in the core region, $r \leq r_c$, when added to the valence potential energy density, will give a perfect cancellation of the ion core potential. The second term in the core region, containing the valence electron density derivative, $df/ds \equiv d\rho/ds$, can be interpreted as a residual kinetic energy density for the valence electrons responding only to the valence potential; here, $s$ represents the increment in energy above the Fermi level of the core electrons. Finally, the second line, with $r > r_c$, is a kinetic energy density for the valence electrons responding to the full outer region potential. The manipulations leading to Eqs. (10), (11) and (12) are all exact, following directly from Eq. (2), the original form of the density.

## V. Unperturbed Closed-Shell Cores and Valence-Only Equations

As separated atoms approach and begin to form a chemical bond, the identity of the individual atoms is lost. We cannot assume that the electron density of the molecule (AB) is a superposition of the constituent atoms (A and B) densities. Therefore, in general,

$$\rho^{AB} \neq \rho^{A} + \rho^{B} . \tag{13}$$

There are, however, exceptions for which the superposition of atomic density retains some validity. When closed-shell, rare-gas atoms interact, the molecular electron density to a good approximation over a wide range of separations is given by a simple superposition of the noninteracting atomic densities. Gordon and Kim calculated the binding energy curves for these rare-gas molecules by assuming superposition of atomic densities obtained from Hartree-Fock calculations for the isolated rare-gas atoms.[8] They then integrated the energy density expressions for a free-electron gas using a Thomas-Fermi form of kinetic energy, as well as an LDA form for the exchange and correlation energies. The results for binding energies and bond lengths at the minimum binding energies were surprisingly good.[8]

In this section, we will borrow from the Gordon and Kim recipe in describing the closed-shell ion cores of interacting constituent atoms. In particular, we will assume that the core electrons are not greatly influenced by the rearrangements of the valence electrons for interaction separations near equilibrium. Under this assumption, we can concentrate on only the valence electron energies. The valence energy is given by the sum of the kinetic and potential energies as

$$\begin{aligned} U_v &= \int d^3 r \left( t_v(\vec{r}) + V_c(r) \rho_v + U_{exc}(\rho_v) \right) + \frac{1}{2} \iint d^3 r d^3 r' \frac{\rho_v(\vec{r}) \rho_v(\vec{r}')}{|\vec{r} - \vec{r}'|} \\ &\equiv T_{eff} + \int d^3 r \left( V_{eff}(r) \rho_v + U_{exc}(\rho_v) \right) + \frac{1}{2} \iint d^3 r d^3 r' \frac{\rho_v(r) \rho_v(r')}{|\vec{r} - \vec{r}'|} \end{aligned} . \tag{14}$$

Here, we have used the LDA form of exchange/correlation, $U_{exc}(\rho_v)$, that depends only on the valence density. For the version in the second line, we used the rewrite of the valence electron kinetic energy density, Eq. (12), and rearranged terms. The residual valence electron kinetic energy in the core region, along with the kinetic energy outside the core region, have been combined into a residual kinetic energy as

$$T_{eff} \equiv \int d^3 r \int_0^{\delta F} ds (s + F_c - V_{eff} - V_v) \frac{d\rho}{ds} \quad . \tag{15}$$

The complete kinetic energy cancellation of the core potential, due to the first term in the first line of Eq. (12), leads to an effective potential, $V_{eff}$, for the closed-shell ion. For an ion with valence, $v$,

$$\begin{aligned} V_{eff}(r) &\equiv 0 & r \leq r_c \\ &\equiv V_c(r) = -\frac{v \cdot e^2}{r} & r > r_c \end{aligned} \quad . \tag{16}$$

The expression for $U_v$ containing $V_{eff}$ suggests that the valence electrons are almost free particles in the inner core region, due to kinetic cancellation, while in the outer region, they respond to the Coulomb potential of the spherical core. Results of this sort, motivated by enforcing orthogonality of the valence and core orbitals, are the underpinnings of the pseudopotential method which has been actively studied for over fifty years.[8,9,10] In our approach, we relied on a direct manipulation of the new DFT version of the valence kinetic energy density, Eq. (11). The details of the ion core potential that were critical for the ionization potential calculations in Section III seem to have disappeared. Actually, the core potential influences $V_{eff}$, $T_{eff}$ and $U_v$ in several ways. First, there is the valence density of states, $d\rho/ds$, that depends on the argument, $(s + F_c - V)$ with $V = V_c + V_v$. Second, there is the core radius, $r_c$, defined by the equation, $F_c - V(r_c) = 0$. Third, the core Fermi level, which is used to fix the number of core electrons, also depends on $V$ through the core density expression. Therefore, the high-potential gradients near the nucleus are still lurking, even though $V_{eff}$ appears to have replaced it with a fairly tame functional form.

We proceed toward minimizing the valence energy by approximating the expression for the residual kinetic energy, $T_{eff}$, converting it to a functional of a valence density. Appendix B develops the valence orbital origins of the approximation. There, by factorizing the orbitals into a radially oscillating function times slowly-varying envelopes, we show that the kinetic energy density contributed by the oscillating factor cancels the core potential in the core region. The smoother valence envelopes are then determined from an effective Hamiltonian, containing a kinetic energy operator and the effective potential, $V_{eff}$. The LSF valence density envelope resulting from this effective Hamiltonian is, up to an energy $s$ above the core Fermi level, given as

$$\rho(s) = \frac{(2m(s + F_c - V_v - V_{eff}))^{3/2}}{3\pi^2 \hbar^3} \quad . \tag{17}$$

With this result, we rearrange the residual kinetic energy term as

$$\begin{aligned} T_{eff} &\equiv \int d^3 r \int_0^{\delta F} ds (s + F_c - V_v - V_{eff}) \frac{d\rho}{ds} = \int d^3 r \int_0^{\delta F} ds \frac{\hbar^2}{2m} (3\pi^2)^{2/3} \rho^{2/3} \frac{d\rho}{ds} \\ &= \int d^3 r \int_0^{\rho_v} d\rho \frac{\hbar^2}{2m} (3\pi^2)^{2/3} \rho^{2/3} = \int d^3 r \frac{3}{5} \frac{\hbar^2}{2m} (3\pi^2)^{2/3} \rho_v^{5/3} \end{aligned} \tag{18}$$

The valence kinetic energy expression from outside the core, the second line of Eq. (11), is carried into the core region as a residual kinetic energy. The final valence energy is

$$U_v = \int d^3 r \frac{3}{5} \frac{\hbar^2}{2m} (3\pi^2)^{2/3} \rho_v^{5/3} + \int d^3 r (V_{eff}(\vec{r}) \rho_v + U_{exc}(\rho_v)) + \frac{e^2}{2} \iint d^3 r d^3 r' \frac{\rho_v(\vec{r}) \rho_v(\vec{r}')}{|\vec{r} - \vec{r}'|} \tag{19}$$

If we minimize this valence energy while enforcing a fixed number of valence electrons, $N_v$, we find

$$U_v - \delta F\, N_v = \int d^3 r \frac{3}{5}\frac{\hbar^2}{2m}(3\pi^2)^{2/3}\rho_v^{5/3} + \int d^3 r (V_{eff}(\vec{r})\rho_v + U_{exc}(\rho_v))$$
$$+ \frac{e^2}{2}\iint d^3 r\, d^3 r' \frac{\rho_v(\vec{r})\rho_v(\vec{r}')}{|\vec{r}-\vec{r}'|} - \delta F \int d^3 r\, \rho_v \quad , \quad (20)$$

in which the Lagrange multiplier, $\delta F$, is used to fix $N_v$. Ultimately, we adjust the increment in Fermi level, $\delta F$, to fix the number of valence electrons.

Eq. (20) is the final embodiment of kinetic cancellation, the valence envelope approximation and the unperturbed core assumption. Most importantly, $U_v - \delta F\, N_v$ is a functional of only a low-spatial frequency valence density. The valence orbital oscillations near the nucleus of each ion core have been removed from the problem. While the core electron density and potential do not appear, their remaining effects are present in the effective potential, $V_{eff}$. Eq. (20) may be applied to molecular and solid-state calculations, where we must include multiple ion cores, or an appropriate lattice of ion cores, as well as the mutual interactions of the cores.

## VI. Estimating the Core Radius

To start a calculation, whether in chemistry or solid-state physics, we need to fix the valence, $v$, as well as estimate the core radius, $r_c$, for each elemental closed-shell ion in the problem. These two parameters then allow us to specify the effective potential, $V_{eff}$, for each constituent closed-shell ion. Essentially all of the proceeding approximations have taken us to Eq. (20) supplemented by the effective potentials, Eq. (16). Unfortunately, while valence is usually well-defined, the core radius, $r_c$, is only defined by the equation $F_c - V_c(r_c) - V_v(r_c) \equiv 0$, so that the core radius depends on the full potential and the core Fermi level. This valence potential, along with the valence density, are the very quantities we want to solve for by minimizing Eq. (20).

Here, we explain one method to obtain an initial estimate of the effective potential for each of the ionized Group III, IV and V elements. We use the ionization potentials calculated in Section III.[10] We proceed by calculating the energy eigenvalue, $\bar{\varepsilon}$, and LSF orbital eigenfunction, $\bar{\psi} \equiv \bar{\phi}/r$, of the s-orbital bound to the effective potential in the radial wave equation (see Eq. (8B)):

$$\frac{d^2 \bar{\phi}}{dr^2} = \left( \frac{2m}{\hbar^2} V_{eff}(r) - \frac{2m}{\hbar^2} \bar{\varepsilon} \right) \bar{\phi} \quad . \tag{21}$$

We choose $v = 3,4,5$ for Group III, IV and V ions respectively, and then make an estimate for the core radius; these two parameters give an initial estimate for $V_{eff}(r)$. Next, we use a predictor-corrector algorithm to solve Eq. (21) and adjust the energy eigenvalue until the eigenfunction approaches zero at large radii. We then make adjustments in the core radius, $r_c$, until the eigenvalue, $\bar{\varepsilon}$, is equal to either the measured or calculated ionization potential, $\varepsilon$, in Table I. Recall that the calculated $\varepsilon$ was obtained by integrating Eq. (9) for the full ion core potential. Figure 1 shows the calculated 5s wave function for the electron bound to the closed shell antimony ion, $Sb^{+5}$, as well as the degenerate solution to Eq. (21); note the exact match for the region $r > R_{ion}$. Continuing with the method based on measured ionization potentials, we calculate the $\tilde{r}_c$ values in Table II.

For all Group III, IV and V elements, the true value of the core radius is close to, but smaller than $\tilde{r}_c$. This is expected, since Eq. (21) only treats the first-valence electron bound to the closed-shell core, while neglecting any readjustments of the core potential, as well as the valence potential due to additional valence electrons. The core radii in the last column of Table II will be used for the band diagram calculations in Section VII.

## VII. Band Structure Calculations[13]

Every electronic structure calculation proceeds by minimizing the total valence energy with respect to the valence density and then solving the resulting equations for the valence density and self-consistent potential. We then insert this potential into a single-electron wave equation. The resulting valence and excited state energy eigenvalues are then compared to data.

When we set the valence density variation of Eq. (20) equal to zero, we find

$$\frac{\hbar^2}{2m}(3\pi^2)^{2/3}\rho_v(\vec{r})^{2/3} + V_{eff}(\vec{r}) + V_{exc}(\vec{r}) + e^2\int d^3r' \frac{\rho_v(\vec{r}')}{|\vec{r}-\vec{r}'|} = \delta F \quad , \tag{22}$$

in which the valence Fermi level increment, $\delta F$, is adjusted to fix the number of valence electrons. The formal solution for the valence density is

$$\rho_v(\vec{r}) = \frac{(2m(\delta F - V_{eff} - V_P - V_{exc}))^{3/2}}{3\pi^2 \hbar^3} \quad , \tag{23a}$$

in which the Poisson potential satisfies

$$\nabla^2 V_P = -4\pi e^2 \rho_v(\vec{r}) \quad , \tag{23b}$$

while the LDA approximation for exchange and correlation gives

$$V_{exc} = -(3/\pi)^{1/3} \cdot e^2 \rho_v(\vec{r})^{1/3} - \frac{e^2}{a_0}(.0311 \cdot \log(a_0 \rho_v(\vec{r})^{1/3}) + .07322) \quad . \tag{23c}$$

Here, we have included a correlation correction, sometimes referred to as the "stupidity energy."[11] For band calculations, the correlation correction can change the band gaps and other features by about ten percent. The correlation energy density whose variation leads to this functional form for potential is discussed in detail in Reference 11. $V_{eff}$ is given in Eq. (16), calculated with appropriate values for valence and the core radius. Once Eqs. (23 a,b,c) are solved to convergence, we have values for both the valence density, $\rho_v(\vec{r})$ and the total potential sensed by the valence electrons,

$$V_T = V_{eff} + V_P + V_{exc}.$$

Equations (23 a,b,c), with kinetic cancellation manifested in the effective potential of the ion cores, are the starting points for electronic structure calculations. All of the derivations and manipulations have led us to this point, and we are now ready to attack some practical problems, such as the band structure of crystalline solids. The key feature of this problem set is the periodic arrangement of the closed-shell ions and neutralizing valence density on a lattice. We will assume that the lattice structure is known, and we will solve for the periodic valence density, as well as the total periodic potential felt by the valence electrons. The band structure follows directly from this periodic potential. In principle, we could include the mutual interactions of the ion cores in Eq. (20), and then find the lattice configuration that minimizes total energy; we will skip this step, and assume that the correct lattice is known.

When lattice periodicity is present in a problem, taking some of the calculation into reciprocal lattice space has advantages.[3,12] For what follows, we will specialize to the zinc-blende lattice with lattice constant, $a$.[12] The set of reciprocal lattice vectors are defined by the property $\vec{g} \cdot \vec{\tau} = Integer \cdot 2\pi$ for any lattice translation vector, $\vec{\tau}$.[3,12] (Reference 12 gives all of the details on the reciprocal lattice vector set used in this work.) When we transform the total lattice valence potential into the reciprocal lattice space, we find the relation

$$V_T(\vec{r}) = V_{eff} + V_P + V_{exc} = \sum_{\vec{g}} (\tilde{V}_{eff} + \tilde{V}_P + \tilde{V}_{exc}) e^{i\vec{g}\cdot\vec{r}} = \sum_{\vec{g}} \tilde{V}_T(\vec{g}) e^{i\vec{g}\cdot\vec{r}} \quad , \quad (24)$$

in which lattice translation symmetry is assured as $V_T(\vec{r}) = V_T(\vec{r} + \vec{\tau})$ for any lattice translation vector, $\vec{\tau}$. The Fourier transform of the effective potential of the ion core lattice is given as

$$\tilde{V}_{eff}(\vec{g}) = \frac{-4\pi e^2}{\Omega(\vec{g} \cdot \vec{g})} (v_c \cos(g\, r_c^c) e^{-i\vec{g}\cdot\vec{s}} + v_a \cos(g\, r_c^a) e^{i\vec{g}\cdot\vec{s}}) \quad . \tag{25}$$

The primitive cell volume is $\Omega$, while $2\cdot\vec{s} = (a/4, a/4, a/4)$ is the basis vector between the cation and anion with valences $(v_c, v_a)$ and core radii $(r_c^c, r_c^a)$ respectively.[3,12] Similarly, the Poisson potential on the lattice is given by the transform of the valence density as

$$\tilde{V}_P(\vec{g}) = \frac{4\pi e^2}{\Omega(\vec{g} \cdot \vec{g})} \tilde{\rho}_v(\vec{g}) \quad . \tag{26}$$

We solve Eqs. (23a) and (23b) iteratively as follows: 1) Using Eq. (23a) with the total potential, $^{old}V_T(\vec{r})$, evaluated in a unit cell of the lattice, we make adjustments in the Fermi level, $\delta F$, until the valence density integrates to $(v_a + v_c)$ valence electrons per unit cell; 2) We calculate the lattice space exchange/correlation potential and then transform it to reciprocal space, $\tilde{V}_{exc}(\vec{g})$; 3) We then transform the valence density to reciprocal space and solve Eqs. (25) and (26) for the sum $\tilde{V}_{eff}(\vec{g}) + \tilde{V}_P(\vec{g})$, while, for charge neutrality, we fix the $\vec{g} = 0$ term to zero; 4) We form the total potential in reciprocal space, $\tilde{V}_T(\vec{g})$, and use linear mixing to estimate the new total potential for the next iteration as $^{new}\tilde{V}_T(\vec{g}) \equiv \tilde{V}_T(\vec{g}) \cdot \beta + {^{old}}\tilde{V}_T(\vec{g}) \cdot (1-\beta)$.[9] We force the new potential to zero beyond a cut-off in reciprocal space and then update the old potential to the new potential, $^{old}\tilde{V}_T(\vec{g}) = {^{new}}\tilde{V}_T(\vec{g})$ ; 5) We transform the potential back to lattice space, $^{old}V_T(\vec{r})$. This completes one iteration. We typically set $\beta = .4$, and we stop the interations when the lattice potential and lattice valence density have converged.

Once the total lattice potential is calculated, we solve a single-electron reciprocal space wave equation for the energy bands. Bloch's theorem for the slowly varying envelopes of the valence states is satisfied by the form

$$\overline{\psi}_n(\vec{r}) = e^{i\vec{k}\cdot\vec{r}} \sum_{\vec{g}} b_n(\vec{g},\vec{k}) e^{i\vec{g}\cdot\vec{r}} \quad , \tag{27}$$

in which $\vec{k}$ is the Bloch momentum.[3] When we neglect the spin-orbit interaction (see Appendix B), the wave equation in reciprocal space is then

$$\left( \frac{\hbar^2}{2m} |\vec{k}+\vec{g}|^2 - E_n(\vec{k}) \right) b_n(\vec{g},\vec{k}) + \sum_{\vec{g}'} \tilde{V}_T(\vec{g}-\vec{g}') b_n(\vec{g}',\vec{k}) = 0 \quad , \tag{28}$$

in which $b_n(\vec{g},\vec{k})$ are the reciprocal space coefficients for the Bloch function and $E_n(\vec{k})$ give the band structure.

We present detailed examples for two technologically important materials: gallium arsenide, GaAs, and silicon, Si. Our band structure results for GaAs, an important laser material, are shown in Fig. 2. For this case, we set the Ga valence to three and As valence to five, using the core radius values given in the last column of Table II. Similar results for Si, with valences set to four, are shown in Fig. 3. For these calculations, we used a cut-off for the potential in reciprocal space at $\vec{g}\cdot\vec{g} \leq 16(2\pi/a)^2$. This cut-off is consistent with calculating only fifteen bands for each material and, therefore, using only the fifteen smallest in magnitude reciprocal lattice vectors in the Bloch functions of Eq. (27); Eq. (28) is then a fifteen-by-fifteen matrix eigenvalue/eigenvector problem with fifteen solutions (bands) at each Bloch momentum .[12] In the band diagrams, we display only six energy bands corresponding to the top three valence bands and three conduction bands. Also, we vary the Bloch momentum from $\vec{k}=0$ to $\vec{k}=\frac{2\pi}{a}(1,0,0)$ for the X-scan and $\vec{k} = 0$ to $\vec{k} = \frac{\pi}{a}(1,1,1)$ for the L-scan in each figure. In addition, the figures show the experimental values for the direct band gap of GaAs, as well as the indirect band gap of Si.[14]

The band structures for the other Group III-V binary and Group IV semiconductors can also be generated, as shown in Fig. 4a-4j. The direct band gap materials, InAs, InP, GaSb and InSb, are found to be direct, while the predicted gaps and other features are fairly accurate.[14] Similarly, the indirect band gap materials, AlAs, AlSb, GaP and AlP, are calculated to be indirect with reasonable accuracy on other band features.[14] In all cases, we expect that slight adjustments in the ion core radii and, of course, the spin-orbit interaction will affect the details. A fairly accurate method for including the spin-orbit interaction in reciprocal lattice space is discussed in Reference 12. When we look to the Group IV band diagrams, we need to point out another sensitivity in the band features. If we reduce the LDA exchange potential to $\kappa = .85$ times the Kohn and Sham version, Eq. (6), then the Group IV band features become more accurate. The results for Si, Ge and Sn were all slightly improved when calculated using this slightly reduced value for the exchange potential. In particular, the band gaps for both Si and Ge were excellent, while the band gap for Sn in the diamond structure, gray tin, was zero, in excellent agreement with the experimental data.[14]

Finally, these results illustrate an extremely important feature: the parameter values that define the effective potential of the ion cores are transferable with reasonable accuracy. For example, the parameters that define the gallium ion core are used in GaP, GaAs, as well as GaSb. That this approximation holds true for all of the other cations and anions leads to extreme efficiencies in all applications. Essentially, all of the chemical bonding properties of any element are, with good accuracy, embodied in its valence and core radius.

## VIII. Conclusions

The electron density approximations as given in Eqs. (2), (3) and (4) retain a high level of accuracy for the ion core potential and should work as well for a large variety of atomic physics problems. The modified form of the density avoids many of the Thomas-Fermi shortcomings, such as infinite density near the nucleus and infinite radius for the neutral atom.[4,5] Our Table I results for the third, fourth and fifth ionization potentials show good agreement with the experimental data, particularly when we include a factor, $\kappa \leq 1$, to lower the strength of the ion core LDA exchange potential. We can motivate this factor by considering the exact exchange contribution to the Hartree-Fock eigenvalue for a plane-wave orbital at an energy above the core Fermi level.[3,4] Finally, we must remember that these calculations are necessarily approximate, as they neglect any influence of the single valence electron on the self-consistent closed-shell ion cores.

Next, we separated core electron densities from valence electron densities by introducing the core Fermi level, $F_c$. Then, when we calculated the valence kinetic energy density, we showed how it separated exactly into two terms, the first of which canceled the potential of the ion cores in the core region, while the second represented the residual kinetic energy driven by the low-spatial-frequency (LSF) valence density. Furthermore, Appendix B demonstrated how these two terms result from factoring the valence orbitals into a rapidly oscillating radial function times slowly varying envelopes. An envelope function approximation then allowed us to write a functional for the total valence energy, in which the effective potential, $V_{eff}$, replaced the ion core potential. These processes of cancellation of the strong ion core potential, kinetic cancellation, and replacing the valence density in the core with a LSF valence density are the most critical and beneficial steps in the entire procedure. That kinetic cancellation might occur in a Thomas-Fermi theory was first suggested in the early days of the development of pseudopotentials.[8] These discussions expanded the traditional Thomas-Fermi kinetic energy density as

$$\frac{3}{5}\frac{\hbar^2}{2m}(3\pi^2)^{2/3}(\rho_c+\rho_v)^{5/3} \approx \frac{3}{5}\frac{\hbar^2}{2m}(3\pi^2)^{2/3}\rho_c^{5/3} + \frac{\hbar^2}{2m}(3\pi^2)^{2/3}\rho_c^{2/3}\rho_v + \ldots$$
$$= \frac{3}{5}\frac{\hbar^2}{2m}(3\pi^2)^{2/3}\rho_c^{5/3} + (F_c - V_c)\rho_v + \ldots \quad ; \quad (29)$$

the second term in this expansion provides kinetic cancellation. This procedure was limited, as it gave no indication of how to include the higher-order terms, was only useful when $(\rho_v/\rho_c)$ was a small quantity and gave no method for fixing the core radius or evaluating the residual valence kinetic energy. The derivations that we gave in Sections IV, V and Appendix B removed these restrictions, yielding the exact core potential cancellation, the residual kinetic energy of the valence envelopes in the core region, and an equation to set the core radius. This kinetic cancellation, based on forming the valence orbitals from the product of a rapidly oscillating radial function and slowly varying envelopes, differs from the standard approach that emphasizes orthogonality of the valence and core electron orbitals.[3,8,9]

In order to calculate with the equations for the LSF valence density, we needed to develop a procedure to set the core radius, $r_c$, for each elemental closed-shell ion in the problem. In Section VI, we discussed an estimate based on measured or calculated ionization potentials; in all cases, this estimate for the core radius provided an upper bound, as shown in Table II. We would certainly like to improve on this estimate to get us nearer to the values in the final column of Table II. Perhaps, in a future method, we might return to the equation, $F_c - V_c(r_c) - V_v(r_c) = 0$, in which the total self-consistent potential, as well as the core Fermi level, appear. This equation defines the boundary of the ion core that is self-consistent with both the core and valence potentials.

In Section VII, we minimized the total valence energy resulting in the valence density and valence potential given in Eqs. (22) and (23). These valence-only equations provide a basis for molecular and solid-state electronic structure calculations. Here, we used them to calculate the band structures resulting from the self-consistent valence density and potential on the zinc-blende and diamond lattices. Our detailed band structure results for GaAs and Si, as well as calculations on other semiconductor materials, indicate that these equations are useful.[14] Also, it appears that the ion cores, as defined by the effective potential of Eq. (16), are approximately transferable among material systems based on Group III, IV and V elements. The parameters that define the gallium ion core can be used in GaP, GaAs, as well as GaSb, and so on for all the closed-shell ion cores. This feature is very important for all implementations of the method.

Why is the new DFT working well in the band structure applications? Accurate band structure calculations have bedeviled the solid-state physics community for eighty years. Numerous approaches and approximations were developed, ranging from the simplest, that ignored the interaction between electrons, to far more complicated schemes that really tried to solve the many-electron problem.[3,8,9] When judged against this history, is the DFT method offered here only a happy accident? Perhaps if the equations worked only for a special case or two, then the method could be discounted. However, we used the new DFT to calculate good band structures for twelve semiconductor materials. Based on these repeated successes in what is a difficult problem set, we feel that Eq. (22) and Eq. (23a,b,c) contain considerable reality. A valence energy expression that depends solely on effective valence potentials, as given in Eq. (16), and low-spatial-frequency valence densities allows for straight-forward, relatively simple calculations that are accurate.

Finally, two comments are necessary: First, in a famous 1962 paper, Teller provided a proof that there could be no chemical binding in the Thomas-Fermi theory, so that separated atoms were at lower total energy than atoms in the molecular state or solid-state.[17] We feel that the Teller proof does not apply to our new DFT, and that accurate chemical binding calculations can be made using Eq. (20) and Eq. (23). Secondly, although this new DFT only applies to systems in which there are equal densities of spin-up and spin-down electrons, it can readily be generalized.[15]

## Appendix A: A Modified Electron Density

We will work a one-dimensional example first. Consider a density for N electrons given in terms of orthonormal orbitals as

$$\rho(x) = 2 \cdot \sum_{n=0}^{M} |\Phi_n(x)|^2 \quad , \tag{1A}$$

in which we have placed a spin-up and spin-down electron into each orbital from the lowest energy, n=0, to the highest occupied level, n=M. We approximate the modulus squared of each orbital using the Wentzel-Kramers-Brillouin method, WKB. In the classically allowed region, with orbital energy greater than the potential, $(E_n - V(x)) \geq 0$, we find

$$|\Phi_n(x)|^2 \approx 2 \left( \sin \int^x dx' \frac{\sqrt{2m(E_n - V(x'))}}{\hbar} \right)^2 \times \frac{N_n}{\sqrt{2m(E_n - V(x))}}$$
$$= (Oscillating\ Factor) \times (Envelope) \quad , \tag{2A}$$

$$\rightarrow \quad \rho(x) \approx 2 \sum_n \frac{N_n}{\sqrt{2m(E_n - V(x))}} \equiv 2 \sum_n \frac{N_n}{p(x, E_n)}$$

in which $N_n$ is a normalization factor for the WKB envelope.[6] In the second line, we have spatially averaged the oscillating $2\langle \sin^2 \rangle \approx 1$ factor in each WKB orbital term; this replaces the modulus squared in (2A) with the envelope factor. We can rewrite the envelope normalization factor for each orbital, $N_n$, if we differentiate the orbital quantization condition with respect to $n$ as

$$\frac{d}{dn} \left[ 2 \int_a^b dx \left(2m(E_n - V(x))\right)^{1/2} = (n + 1/2) \cdot h \right]$$
$$\rightarrow \quad 2m \frac{dE_n}{dn} \int_a^b dx \left(2m(E_n - V(x))\right)^{-1/2} \equiv 2m \frac{dE_n}{dn} \bigg/ N_n = h \tag{3A}$$

The second line can be rearranged to identify the normalization factor as

$N_n = 2m/h \cdot \frac{dE_n}{dn}$. As a check, for a harmonic oscillator potential, $V(x) = \frac{m\omega^2 x^2}{2}$, we can exactly calculate $N_n = \frac{m\omega}{\pi}$ which is equal to $2m/h \cdot \frac{d}{dn}(n\hbar\omega + 1/2)$, so the result fortunately holds in this case. Finally, we rewrite the density based on WKB orbitals as

$$\rho(x) = \sum_n N_n \cdot \frac{2}{p(x, E_n)} = \sum_n 2m/h \cdot \frac{dE_n}{dn} \frac{2}{p(x, E_n)} \approx \int_{E_0}^{F} dE \frac{4m/h}{p(x,E)} + |\Phi_0(x)|^2 \quad . \tag{4A}$$

Here we have replaced the sum over n with an integral supplemented by one-half of the first term at energy $E_0$ in Eq. (1A); these are the lowest-order terms in the Euler-Maclaurin formula.[7] Also, this fixes the lower limit of the integral to $E_0$ and the upper limit is set to the Fermi energy, F; this should be approximately the energy for the highest occupied level. In applications, F is always adjusted to give the correct number of electrons. Finally, we complete the integral over energy to find

$$\rho(x) = 4/h \sqrt{2m(F - V(x))} - 4/h \sqrt{2m(E_0 - V(x))} + |\Phi_0(x)|^2 \quad . \tag{5A}$$

Only the first term, with no lowest-orbital corrections, would be present in the standard one-dimensional Thomas-Fermi result.

The density derivation for a spherically symmetric potential follows a similar path from a slightly more complicated starting point. Consider a density for N electrons given in terms of orthonormal orbitals in spherical coordinates as

$$\rho(r,\theta,\phi) = 2 \cdot \sum_{n=0}^{M} \sum_l \sum_m \left|\Phi_{nl}(r)/r\right|^2 Y_{lm}(\theta,\phi) \cdot Y_{lm}^{*}(\theta,\phi) \quad , \tag{6A}$$

in which $\Phi_{nl}(r)$ is a solution of the radial wave equation and $Y_{lm}(\theta,\phi)$ is a spherical harmonic function. For closed-shell cases, in which all terms $-l \leq m \leq l$ are present, we can use the addition theorem for the spherical harmonics to simplify this to a radial density

$$\rho(r) = 2 \cdot \sum_{n=0}^{M} \sum_{l=0} \left| \Phi_{nl}(r)/r \right|^2 \frac{2l+1}{4\pi} \quad . \tag{7A}$$

We can now follow the development of the one-dimensional example, using the WKB approximation for the solutions to the radial wave equation, averaging over the rapid radial oscillations or, equivalently, invoking a WKB envelope function approximation, and using the normalization obtained from the radial quantization condition. This brings us to the analog of Eq. (4A) as

$$\rho(r) = \frac{2}{4\pi} \cdot \sum_{n=0}^{M} \sum_{l} \frac{dE_{nl}}{dn} \frac{2m}{\hbar^2} \frac{d\beta}{dl} \frac{2m}{h \cdot \sqrt{2m(E_{nl}-V(r)-\beta)}} \quad , \tag{8A}$$

in which we introduced a quantity $\beta \equiv \frac{\hbar^2 l(l+1)}{2mr^2}$ and $\frac{d\beta}{dl} = \frac{\hbar^2(2l+1)}{2mr^2}$; note that this last expression is unchanged if $l(l+1)$ is replaced with Langer's correction, $(l+1/2)^2$.[4,5] Now, as in the one-dimensional case, we approximate the discrete sums with integrals. Here, we have two sets of integral limits to set. For the $\beta$-integration, we use limits at $\beta = 0$ and E-V. Next, for the sum over the principle quantum number, we use the leading terms in the Euler-Maclaurin formula to convert the sum to an integral plus an end-point term from the lowest orbital end-point, $|\Psi_0(r)|^2$. We neglect an end-point contribution from the highest energy orbital, since we will ultimately adjust the Fermi level to obtain the correct total electron number. This leads to

$$\rho(r) \approx \frac{2}{4\pi} \cdot \int_{E_0}^{F} dE \int_{0}^{E-V} d\beta \, \frac{2m}{\hbar^2} \frac{2m}{h \cdot \sqrt{2m(E-V(r)-\beta)}} + |\Psi_0(r)|^2 \quad . \tag{9A}$$

These final integrations give the density

$$\rho(r) = \frac{(2m(F-V(r)))^{3/2}}{3\pi^2 \hbar^3} - \frac{(2m(E_0-V(r)))^{3/2}}{3\pi^2 \hbar^3} + |\Psi_0(r)|^2 \quad . \tag{10A}$$

We discuss the benefits of this formula in the main body of the paper. Primarily, this modified form of the density avoids infinite density near the nucleus and, when exchange is included, infinite radius for the neutral atom.[4] The density expressions, (5A) and (10A), significantly improve the standard Thomas-Fermi density. In each case, the first term on the right-hand-side is the Thomas-Fermi density, but the modifications brought by the second and third terms are significant for atomic physics calculations.

There is an elegant alternative derivation of our new DFT based on operator algebra approximations. The density, Eq. (1A) extended to three dimensions, can be exactly rewritten as a diagonal density matrix element in coordinate space. Using Dirac's bra-ket notation, we write the exact relation

$$\rho(\vec{r}) = 2\langle \vec{r} | \; \theta(F-\hat{H}) - \theta(E_0 - \hat{H}) | \vec{r} \rangle + |\Psi_0(\vec{r})|^2 \tag{11A}$$

in which $\hat{H}$ is the Hamiltonian operator and $\theta$ is the step-function. Guided by the final form of Eq. (10A), we have separated out the lowest energy orbital. We prove this by inserting a complete set of energy orbitals to find

$$\begin{aligned}\rho(\vec{r}) &= 2\langle \vec{r} | \sum_n |\Psi_n\rangle\langle\Psi_n| \left(\theta(F-\hat{H}) - \theta(E_0 - \hat{H})\right) | \vec{r} \rangle + |\Psi_0(\vec{r})|^2 \\ &= 2 \sum_n |\Psi_n(\vec{r})|^2 \left(\theta(F-\varepsilon_n) - \theta(E_0 - \varepsilon_n)\right) + |\Psi_0(\vec{r})|^2\end{aligned} \quad . \tag{12A}$$

Note that the last term, $|\Psi_0(\vec{r})|^2$, occurs because the second step-function is equal to one-half when the orbital energy is exactly equal to the lowest orbital energy, $\varepsilon_n = E_0$. The direct evaluation of the operator expression is difficult because the kinetic and potential energy operators in the Hamiltonian do not commute. One can, following the method of Golden[16], represent the step-functions as an inverse Laplace transform as

$$\theta(F - \hat{H}) - \theta(E_0 - \hat{H}) = \frac{1}{2\pi i} \int_{\gamma-i\infty}^{\gamma+i\infty} dz \frac{e^{zF} - e^{zE_0}}{z} \exp(-z\hat{H}) \qquad \gamma > 0 \quad . \tag{13A}$$

Our entire problem is then reduced to approximating the exponentiated Hamiltonian operator and inverse Laplace transforming. In the lowest-order approximation, we ignore the commutator between the kinetic and potential energy operators. This gives

$$\exp(-z\hat{H}) \approx \exp\left(-z\frac{\hat{P}^2}{2m}\right) \exp(-z\hat{V}) \quad . \tag{14A}$$

We can now directly evaluate Eq. (11A) and Eq. (13A), by placing a complete set of momentum states between the kinetic and potential exponential factors. This gives

$$\begin{aligned}
\rho(\vec{r}) &= 2\langle\vec{r}|\frac{1}{2\pi i}\int_{\gamma-i\infty}^{\gamma+i\infty}dz\frac{e^{zF}-e^{zE_0}}{z}\exp(-z\hat{H})|\vec{r}\rangle + |\Psi_0(\vec{r})|^2 \\
&\approx 2\langle\vec{r}|\frac{1}{2\pi i}\int_{\gamma-i\infty}^{\gamma+i\infty}dz\frac{e^{zF}-e^{zE_0}}{z}\exp\left(-z\frac{\hat{P}^2}{2m}\right)\int d^3p|\vec{p}\rangle\langle\vec{p}|\exp(-z\hat{V})|\vec{r}\rangle + |\Psi_0(\vec{r})|^2 , \\
&= \frac{2}{(2\pi\hbar)^3}\int d^3p\left[\theta\left(F-\frac{p^2}{2m}-V(\vec{r})\right) - \theta\left(E_0-\frac{p^2}{2m}-V(\vec{r})\right)\right] + |\Psi_0(\vec{r})|^2
\end{aligned} \tag{15A}$$

in which we used $\langle\vec{p}|\vec{r}\rangle \equiv e^{-i\vec{p}\cdot\vec{r}}/(2\pi\hbar)^{3/2}$ and the conjugate. The final integrations over spheres in momentum space exactly give Eq. (10A). Remarkably, the WKB approximations, the envelope function approximations and the replacement of orbital sums with integrations over energies are all embodied in the operator approximation in Eq. (14A). Furthermore, with this derivation, we see that the new DFT is not restricted to spherically symmetric problems, while higher-order corrections to the density and the kinetic energy may be systematically derived by improving on the approximation given in Eq. (14A). Although Eq. (11A) leads to a powerful, compact alternative approach to the new DFT, it is perhaps prudent to remember the approximations called out in the derivation leading to Eq. (10A).

**Appendix B: The Valence Envelope Function and Low-Spatial-Frequency Density**

Here, we demonstrate that the kinetic energy contributed by the high spatial frequency radial oscillations in the valence orbitals can exactly cancel the core potential. This supplements Sections IV and V of the paper, providing orbital underpinnings to the kinetic cancellation of the core potential, as well as the residual kinetic energy density. Out to some core radius, $r \leq r_c$, consider writing the valence orbital function as the product of a rapidly oscillating radial function, $\sqrt{2}\sin(u(r))$, and a slowly varying envelope, $\beta_n(\vec{r})$, as

$$\psi_n(\vec{r}) \equiv \sqrt{2}\sin\left(\int_0^r dr' P_c(r')/\hbar\right)\beta_n(\vec{r}) \equiv \sqrt{2}\sin(u(r))\beta_n(\vec{r}) \qquad r \leq r_c \quad , \tag{1B}$$

in which $P_c(r) \equiv \sqrt{2m(-V_c(r))}$ and $du/dr = P_c(r)/\hbar$. We then calculate the kinetic energy of this valence orbital in the core region as

$$\left(\frac{-\hbar^2}{2m}\nabla^2\right)\psi_n = \sqrt{2}\sin(u)\left[\frac{-\hbar^2}{2m}\nabla^2 - V_c\right]\beta_n + \sqrt{2}\cos(u)\left(\frac{-\hbar^2}{2m}\right)\left[\nabla^2 u + 2(\vec{\nabla}u)\cdot\vec{\nabla}\right]\beta_n \quad . \tag{2B}$$

The kinetic energy cancellation in the ion core region can now be developed for these valence orbitals. The Hartree-Fock valence orbital with eigenvalue, $\varepsilon_n$, satisfies the equation

$$\left(\frac{-\hbar^2}{2m}\nabla^2 + V_c + V_v\right)\psi_n = \varepsilon_n \psi_n \quad . \tag{3B}$$

We now substitute in the Eq. (1B) and (2B) results to find

$$\sqrt{2}\sin(u)\left[\frac{-\hbar^2}{2m}\nabla^2 + (-V_c + V_c + V_v)\right]\beta_n +$$
$$\sqrt{2}\cos(u)\left(\frac{-\hbar^2}{2m}\right)\left[\nabla^2 u + 2(\vec{\nabla}u)\cdot\vec{\nabla}\right]\beta_n = \varepsilon_n \sqrt{2}\sin(u)\beta_n \quad . \tag{4B}$$

Equation (4B) is exact in the core region. We can, however, extract an approximate equation that only involves the valence envelope function, $\beta_n(\vec{r})$. Matrix elements between valence orbitals of the form (1B) will always include an additional factor, $\sqrt{2}\sin(u)$ in the core region. Therefore, we multiply Eq. (4B) by this factor. Next, we locally average over radial increments, $\Delta r$, such that $u(r+\Delta r) - u(r) = \pi$, while assuming that the envelope function is essentially constant over the interval. The terms containing the $2\sin^2(u)$ factor dominate, since the average $\langle 2\sin^2(u)\rangle = 1$, while the term containing the $\langle 2\sin(u)\cos(u)\rangle$ factor averages to zero. Therefore, after the incremental averaging, the surviving terms in the core region can be rearranged to

$$\left[\frac{-\hbar^2}{2m}\nabla^2 + V_v\right]\beta_n = \varepsilon_n \beta_n \equiv (s_n + F_c)\beta_n \qquad r \leq r_c \quad . \tag{5B}$$

The valence orbital envelopes are approximated by solutions to this equation; note that only the valence potential survives. This corresponds to $V_{eff} = 0$ in the core region with cancellation of the core potential, while the eigenvalue, $s_n$, is defined as an increment in energy eigenvalue above the Fermi level of the core electrons. Eq. (5B) provides the orbital origins of Eq. (15) and Eq. (16) for $r \leq r_c$. Also, at this same level of approximation, the valence envelopes will conserve the norm out to the core radius as

$$\int_0^{r_c} \left|\sqrt{2}\sin(u(r))\beta_n\right|^2 d^3r \approx \int_0^{r_c} |\beta_n|^2 d^3r \quad . \tag{6B}$$

In the outer region, we smoothly connect to the valence orbital equation

$$\left[\frac{-\hbar^2}{2m}\nabla^2 + V_c + V_v\right]\beta_n = \varepsilon_n \beta_n \equiv (s_n + F_c)\beta_n \qquad r > r_c \quad . \tag{7B}$$

The valence orbital envelopes are then the eigenfunctions of an effective Hamiltonian given by the combination of Eq. (5B) and Eq. (7B) as

$$\hat{H} = \left(\frac{-\hbar^2}{2m}\nabla^2 + V_{eff} + V_v\right) \quad . \tag{8B}$$

The effective potential, $V_{eff}$, is given in Eq. (16). A valence electron density developed from this Hamiltonian will satisfy Eq. (17).

When the spin-orbit interaction is included, Eq. (1B) must be modified by making the valence orbitals Pauli spinors. Also, an additional potential term must be included as

$$V_{SO} = \frac{1}{2mc^2}\frac{1}{r}\frac{dV}{dr}\vec{S}\cdot\vec{L} \quad . \tag{9B}$$

Here, $\vec{S}$ and $\vec{L}$ are the spin and orbital angular momentum operators, and $V$ is the full potential including the ion core. Although kinetic cancellation removes the core potential from the slowly varying envelope equations, the spin-orbit matrix elements must be evaluated using the full-valence orbitals of Eq. (1B). An approximate method for including the spin-orbit interaction in band structure calculations is discussed in Reference 12.

**Acknowledgments**

I want to thank Mr. Michael Tilton for considerable help in both organizing the band diagram results and a careful reading of the final manuscript.

| Element | | I (eV) | K=0.5 | K=1.0 |
|---|---|---|---|---|
| Boron | (B) | 37.93 | 37.15 | 38.30 |
| Aluminum | (Al) | 28.45 | 28.28 | 31.13 |
| Gallium | (Ga) | 30.71 | 32.72 | 38.94 |
| Indium | (In) | 28.02 | 25.66 | 30.46 |
| Thallium | (Tl) | 29.83 | 26.63 | 32.51 |
| Carbon | (C) | 64.49 | 63.45 | 65.02 |
| Silicon | (Si) | 45.14 | 44.63 | 48.01 |
| Germanium | (Ge) | 45.72 | 47.56 | 54.35 |
| Tin | (Sn) | 40.73 | 37.36 | 42.56 |
| Lead | (Pb) | 42.32 | 37.68 | 43.91 |
| Nitrogen | (N) | 97.89 | 96.59 | 98.56 |
| Phosphorus | (P) | 65.02 | 64.14 | 68.03 |
| Arsenide | (As) | 62.63 | 64.35 | 71.70 |
| Antimony | (Sb) | 55.97 | 50.42 | 55.99 |
| Bismuth | (Bi) | 55.97 | 49.48 | 56.40 |

Table I. Measured and calculated ionization potentials.

| Element | | $\tilde{r}_c$ (A) | $r_c$ (A) |
|---|---|---|---|
| *Valence = 3* | | | |
| Aluminum | (Al) | .62 | .61 |
| Gallium | (Ga) | .59 | .56 |
| Indium | (In) | .60 | .60 |
| *Valence = 4* | | | |
| Silicon | (Si) | .57 | .53 |
| Germanium | (Ge) | .56 | .51 |
| Tin | (Sn) | .63 | .57 |
| *Valence = 5* | | | |
| Phosphorus | (P) | .51 | .475 |
| Arsenide | (As) | .53 | .47 |
| Antimony | (Sb) | .61 | .53 |

Table II. Estimated core radii, $\tilde{r}_c$, and core radii, $r_c$, used in band calculations for Group III, IV and V elements.

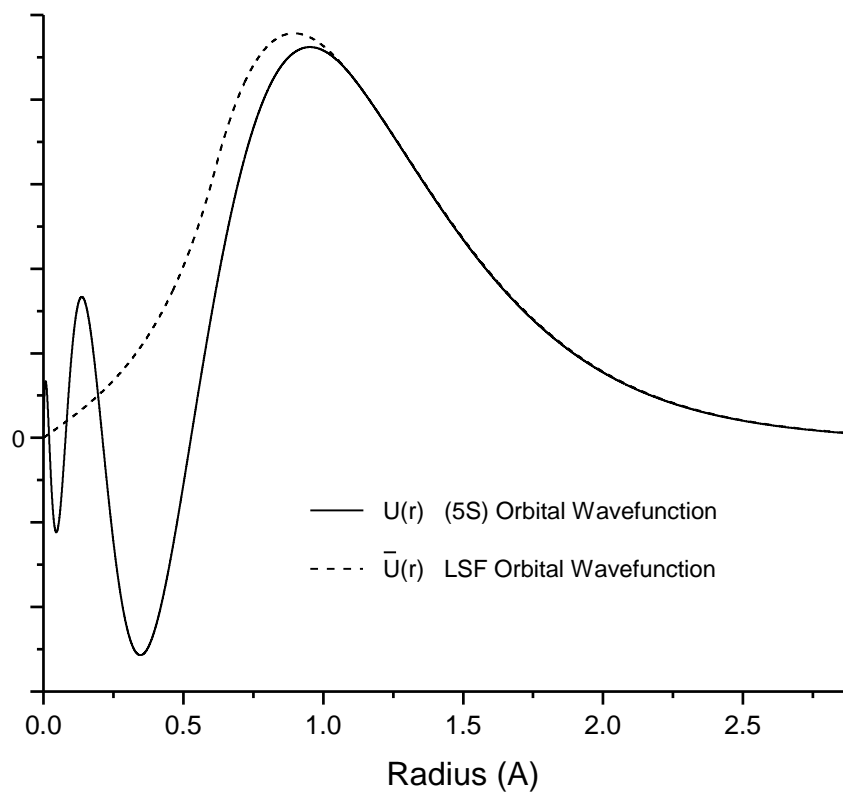

**Fig. (1) Calculated orbital (5S) and degenerate low spatial frequency orbital for a valence electron bound to $Sb^{+5}$.**

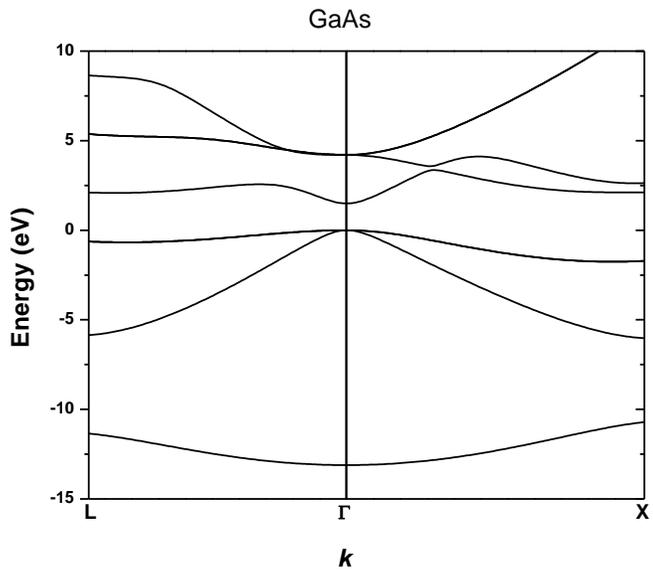 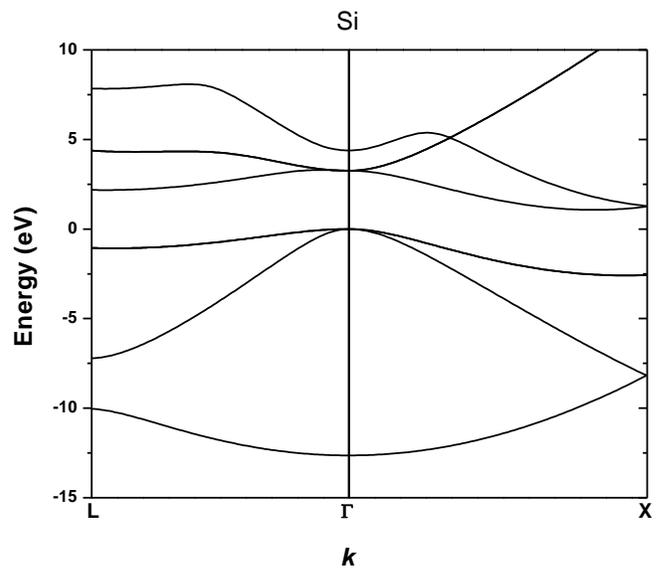

**Fig. (2)** Band Diagram for GaAs.          **Fig. (3)** Band Diagram for Si.

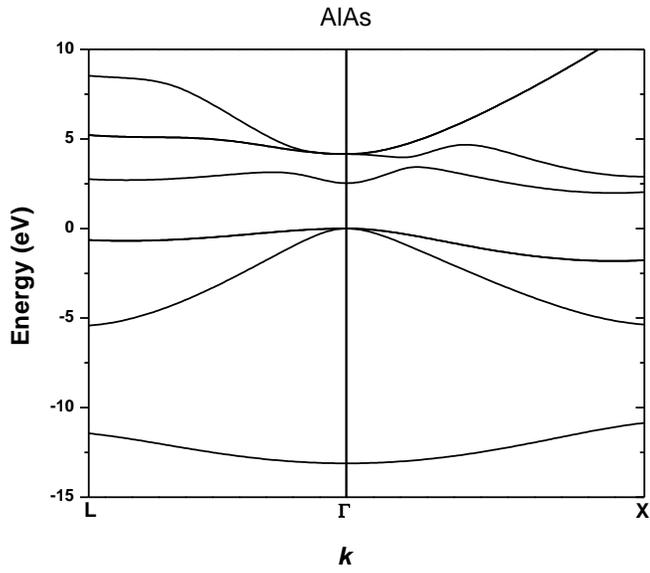

**Fig. (4a) Band Diagram for AlAs.**

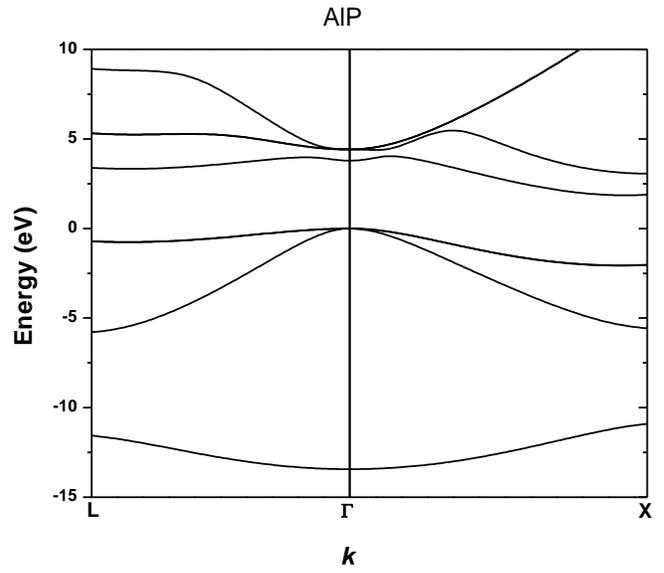

**Fig. (4b) Band Diagram for AlP.**

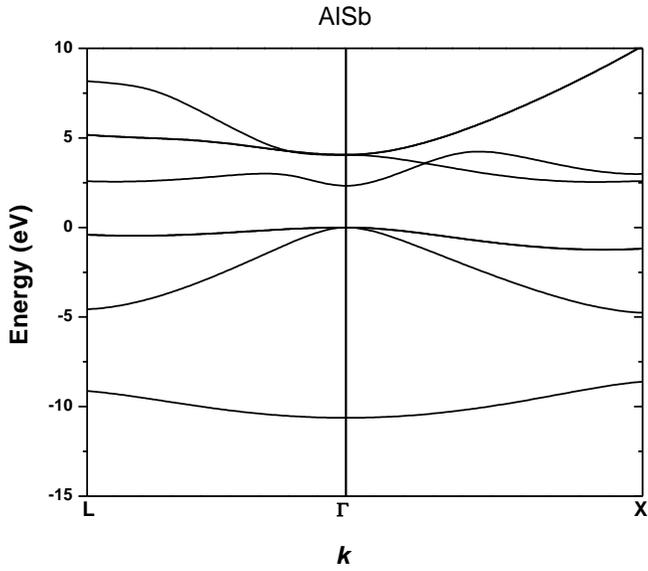

**Fig. (4c) Band Diagram for AlSb.**

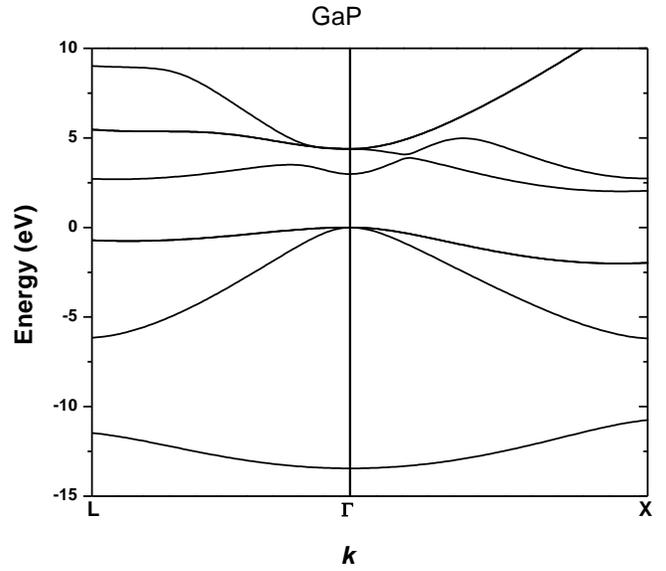

**Fig. (4d) Band Diagram for GaP.**

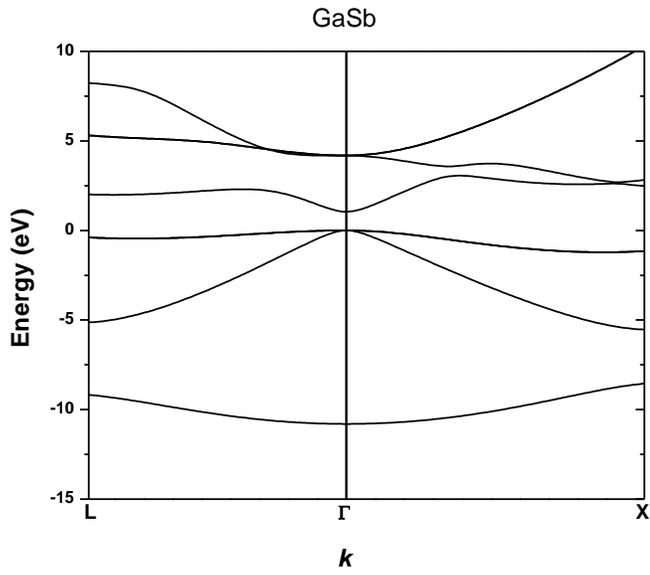

**Fig. (4e) Band Diagram for GaSb.**

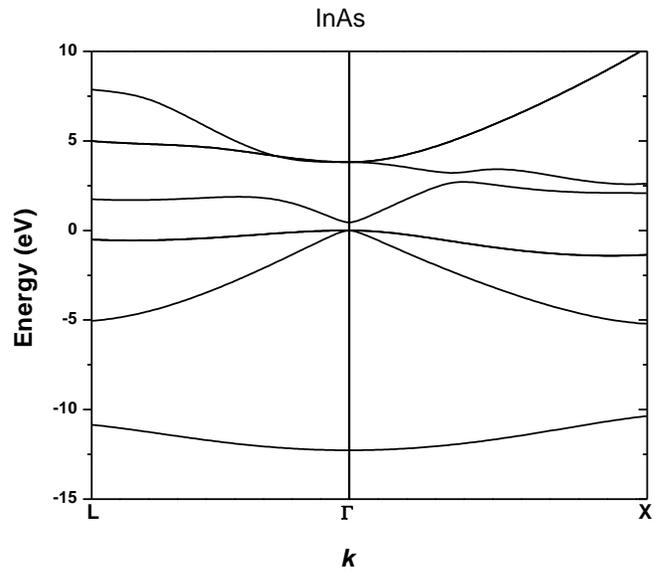

**Fig. (4f) Band Diagram for InAs.**

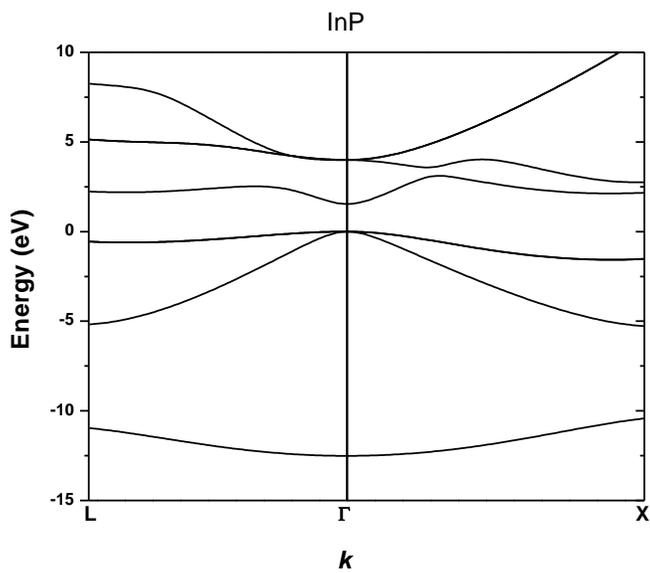

**Fig. (4g) Band Diagram for InP.**

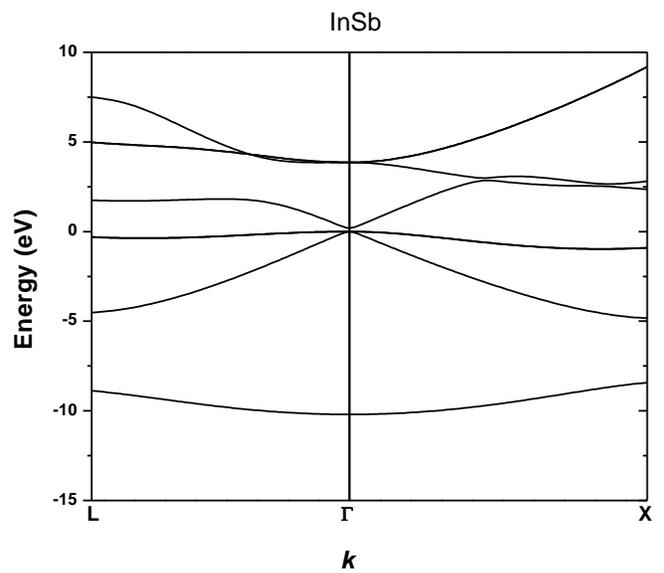

**Fig. (4h) Band Diagram for InSb.**

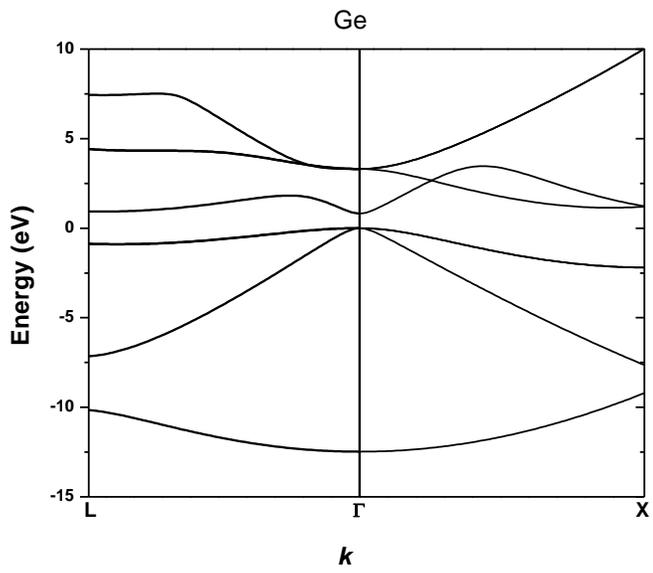 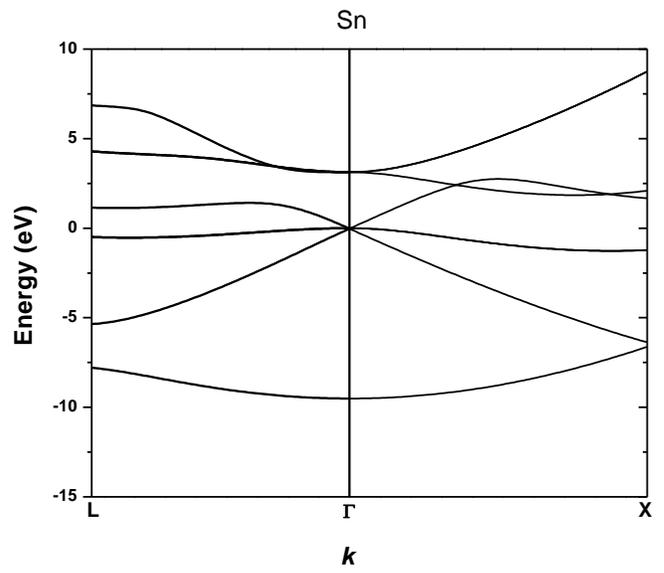

**Fig. (4i) Band Diagram for Ge.**  **Fig. (4j) Band Diagram for Sn.**